# FIDRS: A Novel Framework for Integrated Distributed Reliable Systems

Mehdi Zekriyapanah Gashti

**Department of Computer Engineering**
**Payame Noor University**
**I.R of IRAN**

**Abstract**
In this paper we represent a new framework for integrated distributed and reliable systems. In the proposed framework we have used three parts to increase Satisfaction and Performance of this framework. At first we analyze previous frameworks related to integrated systems, then represent new proposed framework in order to improving previous framework, and we discuss its different phases. Finally we compare the results of simulation of the new framework with previous ones. In FIDRS framework, the technique of heterogeneous distributed data base is used to improve Performance and speed in responding to users and in this way we can improve dependability and reliability of framework simultaneously. In extraction phase of the new framework we have used RMSD algorithm that decreases responding time in big database. Finally by using FDIRS framework we succeeded to increase Efficiency, Performance and reliability of integrated systems and remove some of previous framework's problems.
***Keywords:*** *FIDRS, DSS, Framework, Integrated Systems, Reliability.*

## 1. Introduction

In recent years there was a tendency toward analyzing integrated systems especially about new framework for these systems. Integrated distributed systems are appropriate solutions for organization sources management. Actually integrated systems aren't only doing source planning but also integrate all departments and functions of an organization in an electronic system that can provide the organization with all its needs. Watson described integrated systems as a public term for integrating calculating systems of an organ, Also he defined these systems as a recommended software system that is integrated and can accomplish all needs of an organization such as financial, human, production, selling and marketing sources [1]. Integrated systems were presented includes: concept of IT (Information technology) and business management [2]. These systems simplify organizational calculations and information flow inside it and remove extra data of production processes by integrating. Inter organizational cooperation an organization can keep a uniform and similar view of its processes by using integrated systems [3].The Major idea of integrated systems in using information technology for adopting software applications and organizational processes such as designing, production, buying, marketing and financial aspects [4]. These systematic plans are based on business and are the best way for gaining organizational aims [5]. So these facts tell us that why integrated systems are one of the most important jobs in information technology world and one well-known standards business software in the last decade [2]. Along as performing integrated systems, organizations seek goals such as decreasing cost, decreasing other different parts, having quick access to data, Electronic information communication with users and using new technology for competing with others. They also seek benefits such as quick reply; of information increase in inter organizational cooperation, improvement of relationship with customers, quick decisions, decrease of storage cost, decrease in time of production process and a complete control over business distribution works [5].

Besides the way an integrated system is performed, achieving optimization should be the major aim of organs in performing and operating these systems. Optimization term is the beneficial use of available techniques, human and organizational sources around the integrated system. Technology, organization and users are three important risk factors which have direct influence on processes of performing and operating of integration projects [6]. Re-engineering of commercial activities, specializing of integrated system and users, education should be managed by integrated system project management in order to remove the conflicts between technology, organization and users. If these conflicts are removes, users will use the system move beneficially. And will be more satisfied, so it can lead to optimization in whole organization.

The name of integrated systems has been derived from Material Requirement Planning and Manufacturing Resource planning. Material requirement planning is for calculating material requirement and then the concept of manufacturing resource planning was discussed that new abilities such as sell planning, capacity management and scheduling were added to it. Although the first manufacturing resource planning was discussed as a tool and reasonable step for manufacturing planning but after its practical use by organizations , came to this conclusion that by special Experiments , selling , distribution and human sources can be beneficial for companies and lead to increase in customer's satisfaction . The next step was computer Integrated manufacturing that consisted of technical functions for development of production and production process in an integrated framework. The concept of integrated organizational solution is now called integrated systems [7].

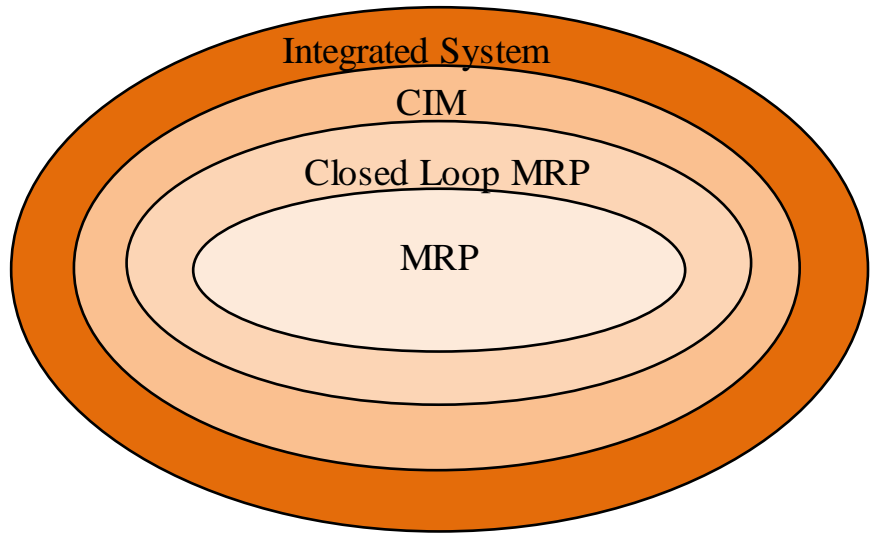

Fig.1 Development of integrated systems

General integrated frameworks are based on manufacturing resource planning functions and technical CA functions. This solution concentrates on producers and manufactures can be expanded easily. And become public designing organizational and integrated data models for the next years were the major focus of CIM (Computer Integrated Manufacturing) Projects. These projects were based on this hypothesis that an integrated data base exists as a central element of information systems and processes modeling when there is an integrated architecture for covering information flow among functions so integrated data models expanded by procedural models growth , in addition to these functions, organization's rules data and organization's application were also covered and today data models and process models are used as reference models in integrated software's [7] (Figure1).

## 2. A Review of Past Frameworks

Integrated systems are findings of IT that are used in a lot of countries around world and are changing and developing rapidly. Although integrated systems have been borne since ago two decades, but they can be considered historically the continuance of a movement that has started from systems of material requirement planning and has developed gradually. Organizational resource planning systems are borne because of funds competitions and increase of their abilities. Planning organizational resources is the attempt in integrating all parts and operations of an organization in a general electronic system that is able to provide all special needs of different parts [8]. As we know, data changes to information after processing and when this information, has a specific order and arrangement changes into knowledge that in this case comes closer to technology. But the point is that data should possess higher security in order to make a possible for us to use it reliably and securely in the electronic world [9]. These issues have been analyzed in details in integrated systems and by integrating them we will be able to increase practicality and beneficiary in a complete information system. Following there is a description of some frameworks for integrated systems.

## 2.1 ERPWKM Framework

Other framework entitled ERPWKM has been proposed in studies which have been done about integrated systems. In this framework the present of accumulation of data in central regions has decreased by WKMSD algorithm and a new technique.

In a comparison between WKMSD and K-Means algorithm authors found some new practical results [10]. WKMSD algorithm presents a new technique to decrease problems and improves the way of information processing in information bank by help of algorithm. This technique possesses bigger productivity and reliability than its previous versions [10]. In a comparison between ERPWKM framework and previous ones which has been done by repetitive simulations and researchers information bank, the present of information accumulation. In its worst mode decreases about 2% and in best mode decreases about 9%, also system's efficiency and reliability increases. Finally the proposed reliable framework was discussed for customers her from inside or outside of organization and inside customers were considered special customers but there were some special customers from outside which were considered as special customers and these special customers could reach their specific services in a shorter time [11].

## 2.2 ERPDRT Framework

ERPDRT framework in integrated systems helps to increase reliability and the ability to be distributed in the system by using simultaneity technique. In ERPDRT different levels of availability are used to increase customer's satisfaction. Inner, outer and middle levels in establishing simultaneity are among different phases of relationship management with customers. Availability level helps us to increase the security of ERPDRT framework in comparison with previous frameworks. When a customer refers to integrated distributed systems which are reliable too, or if they are among customers whom individual information has been registered before some conditions will appear. Condition A is when there is no information about admissive customer in different parts of system so can't have simultaneous access In this condition we can consider more time for system in order to register customers ' individual information and then system can stay in one of following conditions. Condition B is when customer's individual information has been registered in information bank. We can represent a more reliable and proficient service to customers [12].
In ERPDRT framework it has been used of ERPASD algorithm. ERPASD made a new technique which decreased repetitive data in proceeding .For simulating ERPASD algorithm in ERPDRT framework data which was used was from Ministry of education. After simulating the algorithm and checking time of doing the Apriori algorithm and ERPASD algorithm some results were found which showed positive effects during simulating the proposed algorithm and ERPDRT framework and the ERPASD algorithm was more effective than Apriori algorithm. The basic Apriori algorithm had some limitations in searching information banks but ERPASD algorithm presented a new technique which decreased problems more than previous frameworks and by the help of this algorithm , information processing way improved . This work possessed higher proficiency and reliability than its previous versions [13].

## 3. The Proposed FIDRS Framework

After studying about previous frameworks and doing simulating them in different companies and organizations Some problems found, so we're commended a new framework called FIDRS to decrease and remove problems, We used RMSD algorithm in FIDRS framework because of our customers satisfaction and improving our services to them, also we used compound data basis in searching phase of information banks. Following is a description of all parts of FIDRS framework.

### 3.1 Customers Relationship Management phase (CRM)

CRM is a complete way for identifying, attracting and holding customers. Also CRM enables organizations to manage and harmonize relationships with customers through some channels, parts, commercial and geographical ways. CRM is a commercial way that other people, processes and technology to maximize organization's relationships with customers. The true importance of CRM is changing the strategy, practical processes and commercial, business performances in order to attract and hold customers and increase productivity. CRM is a strategy which its goal is understanding predicting and managing organizations and customer's needs.

### 3.1.1 Information Resource Management (IRM)

Organization's information is not only organization's property source but also is a tool for managing other sources and properties of organization. This Value is not practical unless necessary information will be in reach of authorized person in an appropriate time. If data explains summary of relationships between facts, Information will be the definitions which are attributed to data and make a bigger set. IRM is one of tools which defines and describes the process of work and information flow in organs.

### 3.1.2 Sell Configuration & Services System (SCSS)

Often it is believed that sell part and service part should work well together to help organization to improve its conditions. Sell part always regrets that customer service part mentions just minor problems and customer service part believes that sell part throws organization in trouble by giving false and unreal promises. Don't you think that organizations sell their products by these unreal promises? Truth is that for winning in this game sell and service part must play in the same team. Sell configuration and services system is responsible for making a balance between these two parts. This system considers some exceptions for customers to make them satisfied.

### 3.1.3 Strategy

Strategy is a macro program for reaching a unique aim. This term originates from wars planning. CRM strategy defines itself as a major program for gaining the goal protecting and improving it in an organization. Each organization in the deal world should have a strategy for CRM. The most important factor in success of different organizations is customer's satisfaction. Balanced scorecard (BSC) is one of tools for analyzing according to financial norms that organizations use them for studying and estimating customer's satisfaction. If necessary information about customers doesn't exist or it is seldom, we can make two concise groups for customers before performing CRM strategy. If in an organization there is such a condition, before performing CRM strategy it should be a study on customers satisfaction phase.

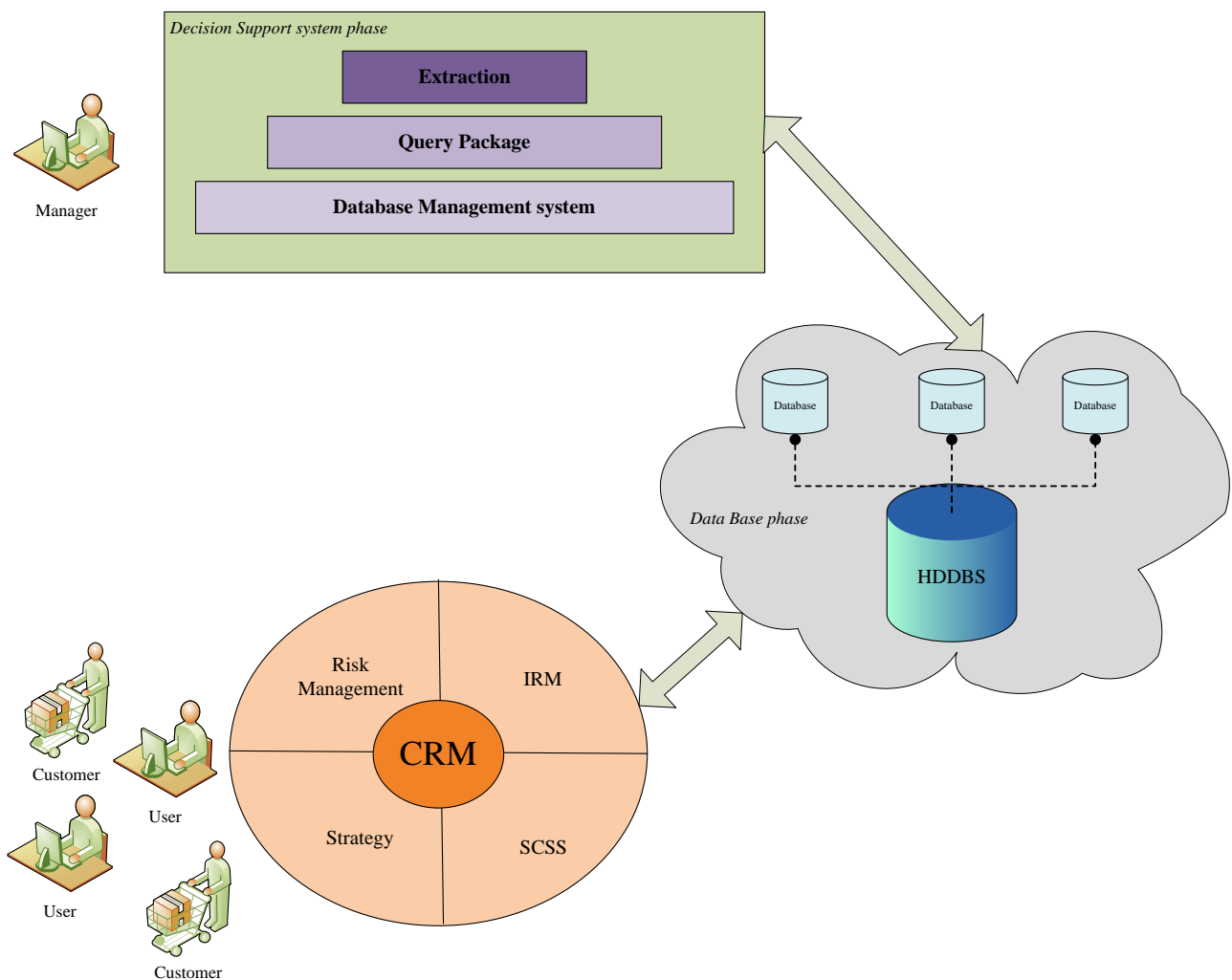

Fig.2 Proposed FIDRS Framework

### 3.1.4 Risk Management

There are different definitions for risk management that all of them have the same meaning and are focused on risk management process such as: Risk management is the process of recognizing, decreasing them down to a reasonable level and finally analyzing its results on the system. Williams and Heinz define risk management so: "Risk management is the process of recognizing, analyzing, and controlling accidental risks that its possible results are damage or unchanging in the condition. Risk management manages risks by controlling them and providing financial damages which have happened beyond efforts for damage control [14]. The most important goal of risk management is to help organization to manage risks better, and the goal of CRM risk management is managing risks which are related to CRM missions such as holding a continuous relationship with organization's customers.

### 3.2 Data Base

Data base is a set of data which is related to minimum unnecessary and extended applications that are independent of electronic and hardware programs. Database is organized in a special way in order to be able to retrieve data if it is needed. They use heterogeneous distributed data base system HDDBS in FIDRS to make an easier relationship among proposed framework phases. In this database they use a standard entitled Gateway Protocol or the same APIs which connect DBMS with Applications which are used to connect different DBMSs such as ODBC, JDBC, etc.

### 3.3 Decision Support System phase (DSS)

Decision making system is an information system which has been designed to help managers in decision making. This system uses data models to solve complex management issues the .major goal in this system is informing manager's about the standard information about their companies and the outside world. This information should consist of time history of processes and organization's outcomes in order to for cast future.

#### 3.3.1. Database Management system (DBMS)

Database has been established by DBMS, and it's up to date by the help of DBMS. Establishing and controlling a database is very hard and complex. DBMS software makes equal all vast and complex files by using a special technique. The recommended framework consists of several databases which some of them are outside of organization. We must add manager's files to database, too DBMS responsible for managing all of these databases.

#### 3.3.2. Query Package

These packages consist of statistical and mathematical models such as linear planning and regression analyzing or special programs for managers or an organization or even an industry. Noticeable models in this field are: strategic and ling run programming, tactical and operational programming and financial programming models.

#### 3.3.3. Extraction

We need a survey or measurement layer for using data and rules in data base. In this layer we used RMSD algorithm to help us to optimize time complexity waiting time and surveying source timing by the help of a new algorithm [15].

Table 2: Result respond time of simulation FIDRS framework

| Number of Request | Respond time in ERWKM (Second) | Respond time in ERPDRT (Second) | Respond time in FIDRS (Second) |
|---|---|---|---|
| 70000 | 0.039 | 0.033 | 0.031 |
| 700000 | 0.571 | 0.601 | 0.528 |
| 3500000 | 1.502 | 1.419 | 1.132 |
| 7000000 | 2.519 | 2.489 | 2.207 |

## 4. Simulation Proposed FIDRS Framework

For simulation the responding time to users and customer's requests it was used of a set of software and hardware equipment's which has been named in table 1.

Table 1: Hardware and Software used for Simulation

| Hardware or Software | Information |
|---|---|
| Processor | Intel 2.4 GHz |
| Memory (RAM) | 2 GB |
| Operating system | Microsoft Windows XP Professional Service Pack 1 |
| Architecture | 32-bit Operating System |

The results of comparing responding times which were gained in previous framework and FIDRS framework and have been presented in table 2.

## 5. Conclusions

In this paper we studied a new framework for integrated systems which a distributed technology was used in it. Results showed that in simulation of FIDRS framework and comparing it with other frameworks that this framework is about 0.02% more practical than other frameworks in some conditions with very low number of requests. The increase of number of requests causes to improve effectiveness present and performance of FIDRS about 12.25% and 08.43% as compared with ERWKM and ERPDRT respectively.

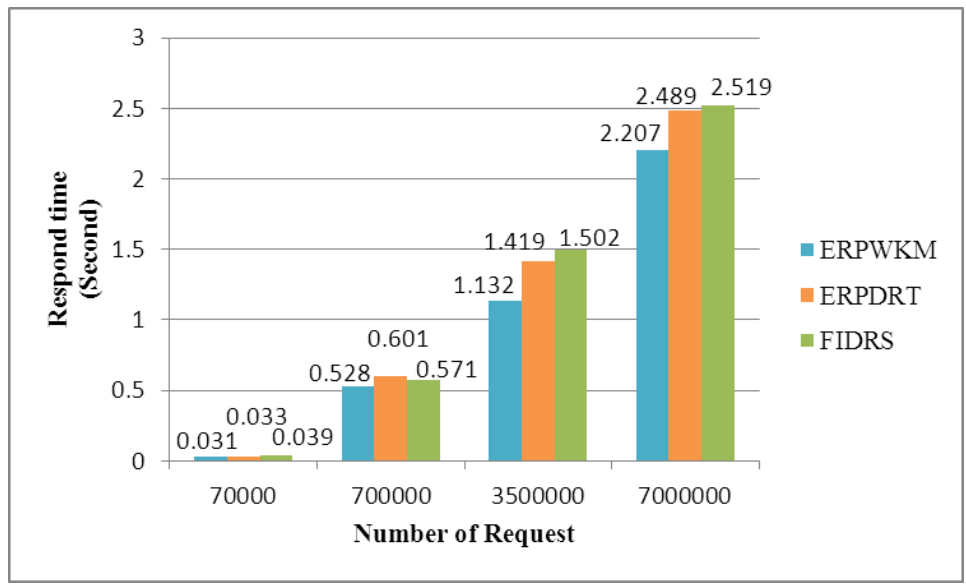

Fig. 3 Compared total processed FIDRS framework and past framework